\documentclass[letter,traditabstract,a4paper]{aa}

\usepackage{natbib}
\bibpunct{(}{)}{;}{a}{}{,} 
\usepackage{graphicx}
\usepackage{subfigure}
\usepackage{amssymb}
\usepackage{txfonts}

\begin{document}

\title{L\MakeLowercase{o}C\MakeLowercase{u}SS: A {\em Herschel} view of obscured star formation\\ in Abell 1835\thanks{Herschel is an ESA space observatory with science instruments provided by European-led Principal Investigator consortia and with important participation from NASA.}}
\author{M. J. Pereira\inst{1}\and C. P. Haines\inst{2} \and G. P. Smith\inst{2} \and E. Egami\inst{1} \and S. M. Moran\inst{3} \and A. Finoguenov\inst{4,5} \and E. Hardegree-Ullman\inst{6}  \and  N. Okabe\inst{7} \and T. Rawle\inst{1} \and M. Rex\inst{1}}
\institute{Steward Observatory, Tucson, Arizona, 85716, USA \email{mpereira@as.arizona.edu}
		\and School of Physics and Astronomy, University of Birmingham, Edgbaston, B15 2TT, England		
		\and Dept of Physics and Astronomy, Johns Hopkins University, 3400 N. Charles Street, Baltimore, MD 21218, USA
		\and Max-Planck-Institut f\"{u}r extraterrestrische Physik, Giessenbachstra\ss e, 85748 Garching, Germany
		\and University of Maryland, Baltimore County, 1000 Hilltop Circle,  Baltimore, MD 21250, USA
		\and Rensselaer Polytechnic Institute (RPI) 110 Eighth Street, Troy, NY 12180, USA
		\and Academia Sinica Institute of Astronomy and Astrophysics, P.O. Box 23-141, 10617 Taipei, Taiwan
		}
\authorrunning{Pereira et al}
\titlerunning{A \emph{Herschel} view of Abell 1835}
\date{Received 31 March 2010 / 
Accepted  6 May 2010}
\abstract
{We present
  \emph{Herschel}/PACS, MMT/Hectospec and XMM-Newton observations of 
  Abell 1835, one of 
  the brightest X-ray clusters on the sky, and the host of a
  strong cool core.  Even though Abell 1835 has a prototypically ``relaxed'' X-ray morphology and no signs of ongoing 
  merger activity in strong- and weak-lensing mass maps, it has a complex
  velocity distribution, suggesting that it is still accreting
  significant amounts of mass in the form of smaller satellite
  systems.  Indeed, we find strong dynamical segregation of
  star-forming dusty galaxies from the optically selected cluster
  population.  Most \emph{Herschel} sources are found close to the
  virial radius of the cluster, and almost a third appear to be
  embedded within a filament feeding the cluster from the SW. We find
  that the most luminous infrared galaxies are likely involved in
  galaxy-galaxy interactions that may have triggered the current
  phase of star formation. }

\maketitle
\markright{}

\section{Introduction}

Galaxy clusters have long been thought of as particularly active sites
of galaxy evolution \citep{Dressler1980}. Early optical studies
suggested that clusters affect stellar populations in infalling
substructure mainly by turning off star formation as galaxies
encounter the dense cluster environment \citep{Lewis2002}. However,
when these same objects were observed in the infrared by IRAS,
ISO and \emph{Spitzer}, a significant fraction of cluster
galaxies was revealed to contain heavily obscured sites of star
formation (see e.g. \citet{Metcalfe2005} and references therein). When
this obscured activity is accounted for, clusters emerge as
environments where star formation is both quenched and induced,
perhaps even at different stages of the same physical process. Some
studies have claimed to observe triggered star formation in strongly
disrupted, merging clusters \citep[e.g.][]{Geach2006} but the intrinsic
scatter in this correlation is large, with some seemingly relaxed
clusters presenting much higher star formation activity than their
actively merging counterparts \citep{Haines2009}.

Many studies in recent years have attempted to separate cluster
samples into two distinct populations: merging clusters vs clusters
that are dynamically relaxed. The global properties of the latter are
expected to be tightly correlated with mass, and therefore directly
relatable to statistical predictions from cosmological models.

While there is evidence that some X-ray cluster properties may in fact
be bimodal, e.g. with respect to the presence or absence of a cool core \citep{Sanderson2009}, cosmological simulations
have shown that even the most relaxed cool core clusters can still be accreting
significant amounts of mass  \citep{Poole2008}, at a mean rate which is approximately
linearly proportional to the mass of the system \citep{McBride2009}.  Generally, only the
most massive of these accretion events (mass ratios $<10:1$) are
expected to disturb the dense cluster core. Massive cool core clusters can thus quietly accrete a significant fraction of their total mass without
  large disturbances to their central ICM,  which makes them arguably ``cleaner" targets for studies of cluster galaxy evolution.

The Local Cluster Substructure Survey
  (LoCuSS\footnote{http://www.sr.bham.ac.uk/locuss}) \emph{Herschel}
key programme \citep{Smith2010}
was designed to probe  dusty 
star-forming galaxies in the infall regions of a sample of 30
galaxy clusters ($z\sim0.2$) spanning a wide range of mass and merging
histories.

Abell 1835 is the most X-ray luminous cluster in this sample ($L_{\rm
  X,bol}=5.32\pm 0.15 \times10^{45}$ ergs/s \citep{Zhang2008}) and the
host of a strong cool core \citep{Peterson2001}. Its
strong- and weak-lensing derived mass distributions \citep{Richard2010,Okabe2010}, and X-ray morphology \citep{Smith2005} are typical of undisturbed (i.e.\ non-merging) cool core
clusters.  For example, the low substructure mass fraction in its core
($f_{\rm sub}=0.13\pm0.01$) suggests that it has grown in mass by
$<10\%$ in the previous few Gyr \citep{Smith2008}.

In this Letter, we focus on a sample of $100\mu{\rm m}$-selected
  spectroscopically confirmed cluster galaxies in Abell\,1835. We use
the Photodetector Array Camera and Spectrometer (PACS)
\citep{Poglitsch2010} on the \emph{Herschel} satellite
\citep{Pilbratt2010} to measure far infrared fluxes and from these we
estimate total infrared luminosities for these objects. We then use
these data to examine the spatial and velocity distribution of
obscured star formation sites in Abell 1835. We assume $H_0= 70$km
s$^{-1}, \Omega_M =0.3,\Omega_\Lambda=0.7$. All virial masses and
radii are stated with respect to the virial overdensity, $\Delta_{\rm
  virial}(z=0.25) =121$ \citep{Bryan1998}.

\section{Observational data}

PACS data were obtained in scan map mode on December 24th 2009, at
both 100 and 160$\mu$m. The details of the observing mode, data
reduction procedure and source extraction and photometry can be found
in \citep{Smith2010}.
The $90\%$ completeness flux limits are 21mJy at 100$\mu$m (86 sources) and 28 mJys at 160$\mu$m (91 sources).

We conducted a wide-field spectroscopic survey of this field with
Hectospec at MMT in February 2009, confirming cluster membership of
$439$ galaxies (Hardegree-Ullman et al., in prep.). In addition, \citet{Czoske2004}
obtained redshifts for 630 cluster members with VIMOS/VLT, bringing the
total number of unique redshifts for this cluster to 851.  This
extensive spectroscopic coverage means that, out of {91
\emph{Herschel} detections with $f_{160\mu m} >28$mJy, 64 have redshifts, of which 33 lie within $\delta v <
  6000$ km/s of the cluster redshift.  The total number of spectroscopically confirmed cluster galaxies with $> 3 \sigma$ detections in both 100$\mu$m and 160$\mu$m bands is  45, and we plot their location in Fig.~1.}

\begin{figure}[]
\begin{center}
\includegraphics[width=0.5\textwidth]{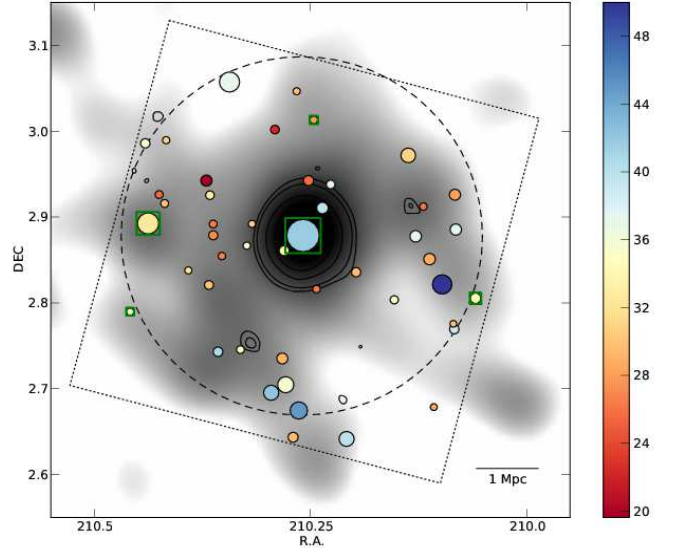}
\caption[Characteristic
Orbits]{\label{fig:orbs1}\footnotesize{Location of \emph{Herschel}
   {detected, spectroscopically confirmed, cluster members on the sky, overlayed on a K-band luminosity
    density image of the cluster produced from all spectroscopically
    confirmed members and corrected for the spatial incompleteness of our
    spectroscopic survey. The black solid contours correspond to 4$\sigma$ wavelet detections of X-ray extended  sources in the XMM data. The dashed circle marks $r_{\rm vir} =
    2.93$ Mpc, the dotted square the limits of the PACS field}. The sizes of the points scale linearly with L$_{\rm
      TIR}$, and their color scales with the temperature of the best
    fit modified blackbody SED, ranging from 20K for the coldest
    galaxies to $\sim 60$K for the hottest. Galaxies identified as AGN
    through emission lines or X-ray emission are marked with a green
    square.}}
\end{center}
\end{figure}

X-ray maps can be used both to separate the AGNs from the starburst
galaxies and to identify galaxy groups around clusters, and we have
analysed the \emph{XMM} {observation ID} 0147330201 to characterize the
X-ray appearance of A1835. We decompose the 0.5-2 keV combined
\emph{XMM} image into unresolved and extended emission, following the
wavelet technique of \citet{Finoguenov2009}, before extracting point
source fluxes at the locations of \emph{Herschel}
sources. Additionally, we make use of previously published wide-field
data from \emph{Spitzer}/MIPS 24$\mu$m and UKIRT/WFCAM J,K
bands\citep{Haines2009}, and Subaru/Suprime-CAM \citep{Okabe2010}.

\section{Results}

\subsection{Infrared luminosities, temperatures and SFRs}

\begin{figure}[]
\begin{center}
\includegraphics[width=0.5\textwidth]{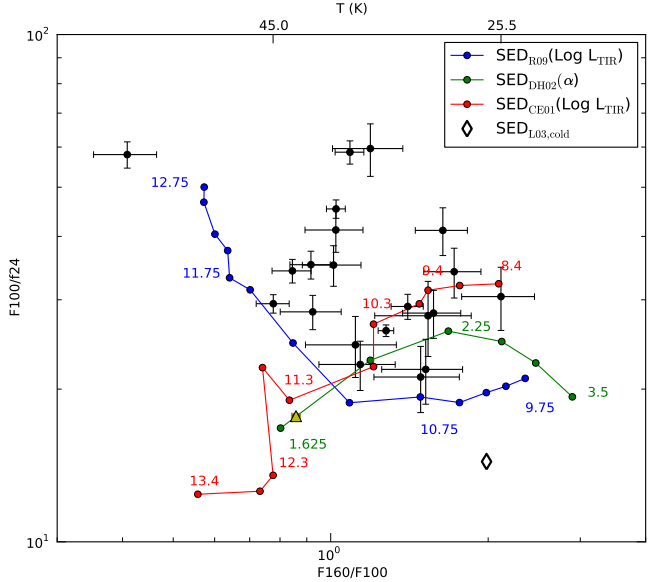}
\caption[Characteristic
Orbits]{\label{fig:orbs1}\footnotesize{160$\mu$m/100$\mu$m vs
    100$\mu$m/24$\mu$m colors, compared with synthetic colors derived
    from SED templates in the literature. The 100$\mu$m/24$\mu$m colors of the cluster galaxies appear significantly reddened compared to the templates. The BCG
    (yellow triangle) has the lowest 100$\mu$m/24$\mu$m color of the
    sample, and is well fit by a Dale \& Helou starburst SED with
    $\alpha \approx 1.65$. Error bars are based on
    statistical noise, and do not include the overall PACS calibration
    error. {The stated PACS calibration uncertainty of $< 10\%$ at 100$\mu$m is nevertheless not large enough to reconcile the data with the template colors.}}}
\end{center}
\end{figure}

At the redshift of the cluster, $z=0.25$, the PACS 100 and 160$\mu$m
bands are expected to straddle the peak of the spectral energy
distribution for sources with dust temperatures of T $\sim
30$K. The 160$\mu$m/100$\mu$m color should thus be sensitive to dust
temperature differences, with redder objects having cooler dust
components. We plot observed 160$\mu$m/100$\mu$m vs 100$\mu$m/24$\mu$m
colors for the sources in our cluster in Fig.~2, and overplot the synthetic
colors from commonly used templates from the literature:
\citealt{Rieke2009} (R09), \citealt{Dale2002} (DH02), \citealt{Lagache2003} (L03), \citealt{Chary2001} (CE01). All the SED templates are poor representations of these galaxies, which appear too red in their 100$\mu$m/24$\mu$m colors. {The CE01 templates appear to fit the colors of $\sim 1/2$ of the sample, but with templates corresponding to total infrared luminosities that are one to two orders of magnitude fainter than the actual measured luminosities} -- see \citet{Smith2010} and \citet{Rawle2010} 
 for a more extended discussion of this $100\mu{\rm
    m}$ excess.

{We estimate total infrared luminosities, $L_{\rm TIR}$, for these galaxies in two ways: first, we fit the CE01 templates (redshifted to the cluster rest-frame) to the 24, 100 and 160 $\mu$m emission and integrate the best-fit spectrum, normalized to the observed fluxes, from rest-frame 3-1100$\mu$m.} In order to obtain estimates for the dust temperatures in these systems, we also fit a modified blackbody spectrum to the 100$\mu$m and 160$\mu$m emission: $f_\nu \propto [ 1 -
\exp(- \tau_\nu) ] B_\nu$, where $B_\nu$ is the Planck
function,
and
$\tau_\nu$ is the frequency-dependent optical depth,
${\nu/\nu_0}^\beta$. We assume $\beta=1.5$ as the emissivity power law
index, and {$\nu_0 = 3$ THz} as the frequency at which the emission
becomes optically thick, but note that the integrated
luminosity depends very weakly on the values assumed for $\beta \in
[0.7-2]$ and $\nu_0$, while the best fit temperature varies at most by
$20\%$. {The $L_{\rm TIR}$ derived from the blackbody fits are systematically lower than the ones obtained from the best-fit templates by a factor of $\sim 0.6-0.8$, since they underestimate the mid-IR contribution to the total IR flux. The cluster members range in L$_{\rm TIR}$ from $4-79\times10^{10} L_{\sun}$, and all but one have dust temperatures in the range $20 - 45$K. }

The BCG is much brighter than the rest of the galaxies in the cluster,
with fluxes of $255.6\pm3.9$ mJy (160$\mu$m) and  297.8$\pm$3.1 mJy
(100$\mu$m) and an estimated $L_{\rm TIR}$ {of  $7.9 \times10^{11}$,}
consistent with previous Spitzer/MIPS observations of Abell 1835
\citep{Egami2006}.  We derive star formation rates from the combined H$_{\alpha}$ (from
our spectroscopic measurements) and L$_{\rm TIR}$ emission using the
following relation from \citet{Kennicutt2009}: ${\rm SFR [Kroupa]}
(M_\odot \,yr^{-1}) = 5.5 \times 10^{-42} [L(H\alpha)_{obs} +
0.002\,L_{\rm TIR}] \ ({\rm ergs\,s^{-1}})$. This yields SFRs in the
range: {2--20 }$M_\odot \, yr^{-1}$ for the general cluster
  population.
  The BCG's high luminosity and special location at the center of a massive cool core separate it from the rest of the cluster population and we therefore exclude it from the subsequent analysis. 
  
We find that, of all galaxies with far-infrared emission detected by
\emph{Herschel}, only 2 have X-ray counterparts down to L$_X \approx 4\times10^{41}$ ergs s$^{-1}$, which is below the classical $10^{42}$ ergs s$^{-1}$ threshold for AGN definition. Only four galaxies (including the couple of X-ray point
sources) in the sample are spectroscopically classified AGN according
to their position in the classic BPT diagram \citep{Baldwin1981},
defined by the emission line flux ratios of [NII] 6587/H$_\alpha$ versus
[OIII] 5007/H$_\beta$. We thus conclude that the bulk of the infrared
luminous population is being heated by star formation and not an
active nucleus, and we exclude the 4 likely AGN from our analysis.

\subsection{Probing the environments of obscured star formation}

It is apparent from Fig.~1 that, apart from the BCG, there are few
luminous infrared galaxies in the central regions of the cluster. In
fact, the radial surface density profiles in Fig.~3 show that
most of the infrared luminosity is concentrated at large ($r \sim 2.5
$Mpc) radial distances, which is not the case for the K-band
luminosity density which, as an approximate tracer of stellar mass,
declines smoothly with radius.

\begin{figure}[]
\begin{center}
\includegraphics[width=0.5\textwidth]{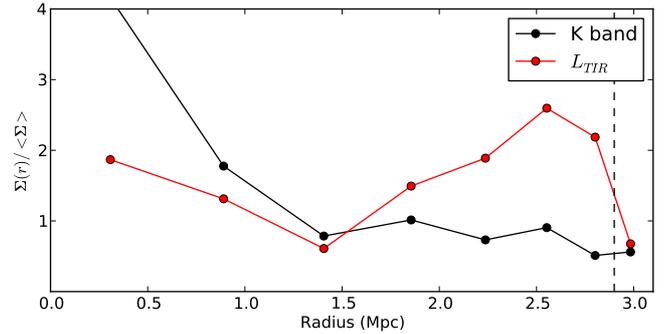}
\caption[Characteristic Orbits]{\label{fig:orbs1}\footnotesize{Radial
    profile of K-band (black) and total infrared (red) luminosity
    surface density relative to the average surface density {in concentric annuli within r}
    $<$3 Mpc (excluding the BCG). The dashed vertical line marks
    r$_{\rm vir}$. The K-band luminosity density is an approximate
    tracer of stellar mass, and declines continuously with radius,
    whereas the infrared luminosity peaks close to the virial radius
    of the cluster.}}
\end{center}
\end{figure}

Upon closer inspection, however, it is clear that this overdensity at
the virial radius does not occur isotropically around the cluster, but
rather peaks along specific directions. The velocity distribution of
K-band selected cluster galaxies (black line in  Fig.~4) is
distinctly non-Gaussian, and shows signs of significant
substructure. The most luminous infrared sources, located toward the
south of the cluster, are dynamically segregated from the rest of the
cluster population, clustering around an elongated structure in the
foreground ($\delta v \sim 2000$ km/s) of the cluster, which is also
traced out in the K-band luminosity density maps (see {Fig.~1}). A
gradient is observed in the velocities projected along the axis of
elongation, where the relative velocity with respect to the cluster
center progressively decreases as the inner region of the cluster is
approached.

\begin{figure}[]
\begin{center}
\includegraphics[width=0.5\textwidth]{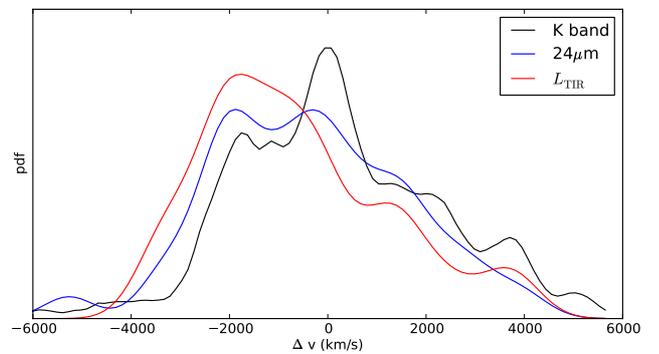}
\caption[Characteristic
Orbits]{\label{fig:orbs1}\footnotesize{Redshift probability density
    function for galaxy populations observed at different
    wavelengths. The density function is weighted by flux/luminosity,
    and corrected for the incompleteness in our spectroscopic
    survey. K-band: black, MIPS 24$\mu$m:blue, $L_{\rm TIR}$: red }}
\end{center}
\end{figure}

Figure~5 shows Subaru  $i'$-band cutouts of the most
luminous infrared galaxies, in decreasing order of $L_{\rm TIR}$. Many
galaxies appear to have undergone recent interactions - $\sim$ half of
them have close companions, and half present distorted
morphologies indicative of recent merging. We can attempt to quantify
the importance of mergers in our small sample by comparing the
fraction, $f_c$, of \emph{Herschel} sources with close ($\Delta r <
50$ kpc and $\Delta v < 300$ km/s) companions to that of a population
of non-IR luminous cluster galaxies that has been matched, both in
K-band luminosity and radial distance from the cluster center, to the
IR luminous sample (excluding in both samples all identified
AGNs). We find that, while the total sample of IR luminous galaxies is
indistinguishable from the matched sample of IR faint galaxies ($f_c
\sim 12^{+8}_{-5}\%$ vs $10^{+2}_{-2}\%$, respectively), the IR bright
($\log_{10} L_{\rm TIR} > 10.8$) galaxies in Abell 1835 are marginally
more likely to have a close neighbour ($f_c \sim 30^{+21}_{-16}\% $ vs
$9^{+3}_{-2}\%$ in the respective matched sample).

\begin{figure}[]
\begin{center}
\includegraphics[width=0.5\textwidth]{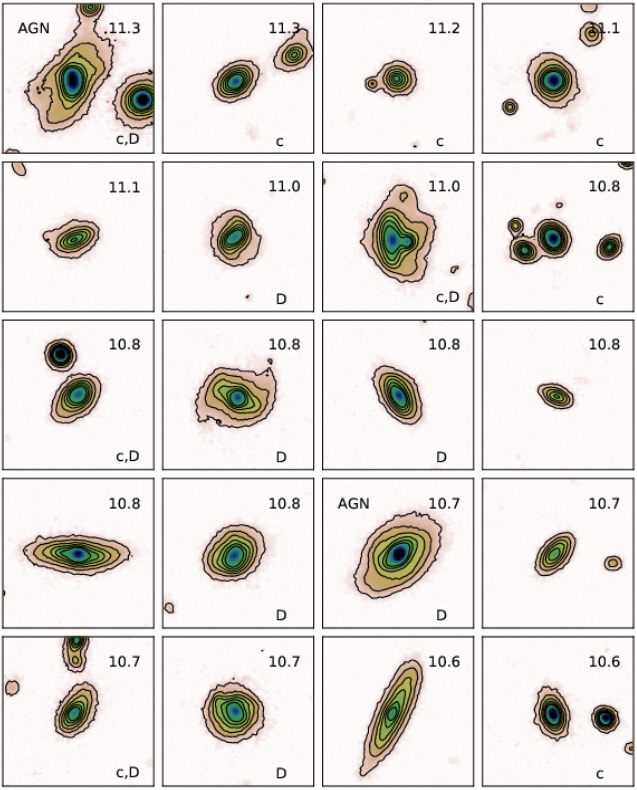}
\caption[Characteristic Orbits]{\label{fig:orbs1}\footnotesize{Subaru
    i$^\prime$ band $\sim 60\times60$ kpc image cutouts centered on
    the IR brightest cluster galaxies, in decreasing order of
    $\log_{10} L_{\rm TIR}$ (upper right corner). A significant fraction has close companions (c), and half show low surface
    brightness tidal features, or asymmetric morphologies indicative
    of a recent interaction (D)}}
\end{center}
\end{figure}

\section{Summary and discussion}

Star formation can be triggered by several different physical
  processes prevalent in clusters and their infall regions; disentangling these effects is one of the main aims of the LoCuSS \emph{Herschel} key programme. The most promising
candidates, ICM ram pressure, galaxy mergers, and galaxy harassment
can all perturb the ISM to produce a new generation of stars, but the
relative importance of these processes is a strong function of
clustercentric radius (c.f.\ Fig.~1 in \citet{Smith2010}).
The location of the peak of IR luminosity density just inside the
virial radius of Abell 1835 suggests that merging may be
driving the bulk of the star formation in these galaxies. This
interpretation is corroborated by the large fraction of disturbed
morphologies in the infrared luminous sample, and also by our
preliminary analysis of their local environment: when compared to a
matched sample of galaxies with no detectable infrared emission,
infrared bright galaxies are more likely to have close neighbours, although the larger sample of 30 LoCuSS clusters should
be used to confirm this correlation.

The $L_{\rm TIR}$ density profile for Abell 1835 is not spherically
symmetric: the brightest galaxies are clustered together in an
elongated structure towards the south of the cluster that is
dynamically segregated from the rest of the cluster. The structure's
K-band morphology and velocity gradient toward the cluster core
suggests that these galaxies may be embedded within a filament that is
still currently feeding the cluster with new, actively star forming
galaxies. 

\emph{XMM} imaging of this field reveals the
presence of two groups in the outskirts of Abell 1835 (see contours in
Fig.~1), one of which appears to be embedded in the filament
that is also feeding the bright LIRGs into the cluster. The group is
dominated by a large {(L$_K\approx 2\times10^{12}$L$_\odot, z=0.245$)} elliptical galaxy
with no detectable \emph{Herschel} emission. 

The two groups have X-ray luminosities of $4.6\pm0.6$ and $7.5\pm0.8$
$\times10^{42}$ ergs s$^{-1}$ (0.1-2.4 keV), from which we estimate
masses, using the weak lensing calibration of the $L_X-M$ relation of
\citet{Leauthaud2010}, of $M_{200} = 4.3$ and $5.8 \times10^{13}
M_\odot$ respectively. Together these amount to $\gtrsim$ 1/10th of
the mass of the central cluster, reinforcing the view that
  Abell\,1835 continues to grow by feeding on smaller systems from the surrounding large scale structure.

\acknowledgements{We thank our colleagues in the LoCuSS collaboration for stimulating discussions. This work is based in part on observations made with
  \emph{Herschel}, a European Space Agency Cornerstone Mission with
  significant participation by NASA.  Support for this work was
  provided by NASA through an award issued by JPL/Caltech. CPH thanks STFC for support. GPS is supported by the Royal Society }

\bibliographystyle{aa}     

\bibliography{refs}

\end{document}